\documentclass[aps,pre, twocolumn, groupedaddress, 9pt]{revtex4-1}
\usepackage{amsmath}

\usepackage{bbding} 
\usepackage{amssymb}
\usepackage{amsfonts}
\usepackage{dcolumn}
\usepackage[export]{adjustbox} 
\usepackage[framemethod=TikZ]{mdframed}
\usepackage{xcolor}
\usepackage{setspace}
\usepackage{caption}
\usepackage{subcaption}
\usepackage{lipsum}
\usepackage{tcolorbox} 
\usepackage{mathtools} 
\usepackage{tgpagella}
\usepackage{titlesec}
 \usepackage{libertine } \usepackage[libertine]{newtxmath} 

\DeclareMathOperator{\Tr}{Tr}

\newcommand{\refew}[1]{Eq.\eqref{eq:#1}}
\newcommand{\rfw}[1]{Eq.\eqref{eq:#1}}
\newcommand{\reffig}[1]{Figure \ref{fig:#1}}
\newcommand{\refsec}[1]{Sec. \ref{sec:#1}}

\newcommand{\mm}[0]{\nonumber \\}
\newcommand{\ve}[0]{\varepsilon}

\titlespacing*{\subsection}
{0pt}{2.5ex plus 1ex minus .2ex}{2.5ex plus .2ex}

\begin{document}


\title{Permutation glass }



\author{Mobolaji Williams} 
\affiliation{%
Department of Physics, Harvard University, Cambridge, MA 02138, USA 
}%
\email{williams.mobolaji@g.harvard.edu}

\date{December 9, 2019}

\begin{abstract}

The field of disordered systems provides many simple models in which the competing influences of thermal and non-thermal disorder lead to new phases and non-trivial thermal behavior of order parameters. In this paper, we add a model to the subject by considering a system where the state space consists of various orderings of a list. As in spin glasses, the disorder of such ``permutation glasses" arises from a parameter in the Hamiltonian being drawn from a distribution of possible values, thus allowing nominally ``incorrect orderings" to have lower energies than ``correct orderings" in the space of permutations. We analyze a Gaussian, uniform, and symmetric Bernoulli distribution of energy costs, and, by employing Jensen's inequality, derive a general condition requiring the permutation glass to always transition to the correctly ordered state at a temperature lower than that of the non-disordered system, provided that this correctly ordered state is accessible. We in turn find that in order for the correctly ordered state to be accessible, the probability that an incorrectly-ordered component is energetically favored must be less than the inverse of the number of components in the system. We show that all of these results are consistent with a replica symmetric ansatz of the system and argue that there is no permutation glass phase characterized by replica symmetry breaking, but there is glassy behavior represented by a residual entropy at zero temperature. We conclude by discussing an apparent duality between permutation glasses and fermion gases.

\end{abstract}


\maketitle


\section{INTRODUCTION}

\setlength{\parskip}{0pt}
In statistical physics, spin glasses exist as archetypical models of disorder due both to their solubility and to the fact that they lend intuition to systems outside of physics which nonetheless exhibit properties common to many spin glasses. Soon after the first spin glass models were solved, physicists sought to apply the lessons of frustration, quenched disorder, and multiple equilibria to biological systems like neural networks \cite{hopfield1982neural, hopfield1984neurons} and proteins \cite{bryngelson1987spin}. But because biological systems integrate structure, function, and dynamics in ways not mirrored by any canonical model of physics, the utility of these spin glass models existed not in providing detailed predictions about biology but in supplying a quantitative framework in which to develop new ways of understanding and describing biological problems \cite{stein1992spin} . 

In a previous work \cite{williams2017statistical}, we moved in the opposite direction: Rather than using our understanding of physics to develop new questions about biology, we used a biological question to motivate the inquiry into a physical system. Motivated by a computational examination of the protein design problem \cite{shakhnovich1998protein}, we considered a statistical physics model of permutations in which the state space was isomorphic to the symmetric group. Significantly, the model's motivation came, not from a physical system, but from a Monte Carlo study of a problem of biochemistry, and to establish intuition for it we considered lattice models where the energy costs were uniform across the system. But even with this simple assumption, the resulting permutation model had interesting thermal behavior because the non-factorizable nature of the state space conferred entropic disorder to a system which was nominally non-interacting. Thus the system exhibited thermal transitions among units which were coupled through state space even though they were not coupled in the Hamiltonian. 

In this paper, we connect our study of the statistical physics of the symmetric group to disordered systems by considering the properties of a system with a state space of permutations and a quenched distribution of energy parameters. Given the unique nature of the state space and the solubility of the non-disordered analog, such a permutation glass \cite{Note1} offers opportunities to explore the relationship between equilibria and disorder in simple exactly soluble physical systems. 

In Sec. \ref{sec:two} of this paper, we discuss the original permutation model, provide schematic depictions of the systems to which it applies, and derive equations defining the thermal equilibrium of the permutation glass. In Sec. \ref{sec:three} we consider the permutation glass for various distributions of energy costs and derive transition temperatures for each case noting their overall consistency with the general result that $k_BT_c \leq \overline{\lambda}/\ln N$, where $\overline{\lambda}$ is the mean of the energy-cost distribution and $N$ is the number of components in the system. That is, the transition temperature of a permutation glass is always less than the transition temperature of the non-disordered system with energy cost given by $\overline{\lambda}$. In Sec. \ref{sec:sim} we compare the computed transition temperatures and a general expression for the order parameter of the permutation glass to results from simulations. In Sec. \ref{sec:understand} we derive a distribution-independent result requiring that the ``completely correct" state can only be a thermodynamic equilibrium of the system if $P_{\lambda<0} <1/N$ where $P_{\lambda<0}$ is the probability that an incorrectly-ordered component is energetically favored. In Sec. \ref{sec:four} we use our derived results to define a glassy regime (which although not representing a distinct phase) cannot be found in the non-disordered system. In Sec. \ref{sec:five} we conclude by discussing ways to extend this simple model of a permutation glass to more complicated models that could exhibit replica symmetry breaking, and we present an analogy between this system and a system of fermions.

\section{EQUILIBRIUM OF PERMUTATION GLASS \label{sec:two}}

\begin{figure}[b]
\centering
\includegraphics[width=.85\linewidth]{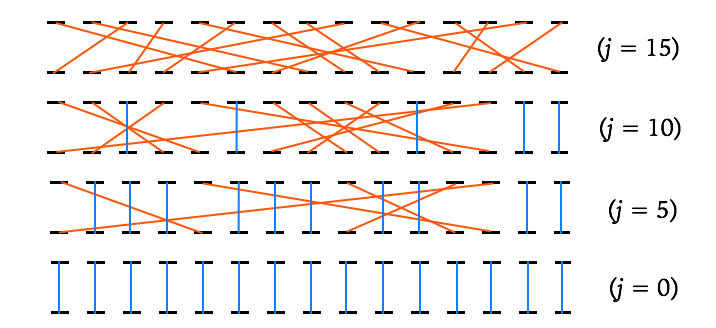}
	\caption{The permutation graph depiction of four microstates in a permutation system with $N=15$. In each graph, $j$ is equivalent to the number of diagonal lines in the permutation graph. The number of ``correct" connections are shown as vertical lines. If we associate a Boltzmann factor $e^{-\beta\lambda_k}$ with each bottom slot $k$ which is not connected to its corresponding top slot, multiply all Boltzmann factors for a graph, and then sum over all possible permutation graphs weighted by their net Boltzmann factor, we obtain \refew{partfunc1}.} 
	\label{fig:permgraph}
\end{figure}

In the statistical physics of permutations presented in \cite{williams2017statistical}, we considered a state space defined by a list of $N$ unique components $(\omega_1, \omega_2, \ldots, \omega_N)$. Taking the states of the system to be the various $N!$ orderings of the components, and defining the zero-energy state as the state where the components are in the order $(\omega_1, \omega_2, \ldots, \omega_N)$, we can postulate a simple Hamiltonian in which there is an energy cost $\lambda_k$ for each state where $\omega_k$ is not in the position given by its zero-energy ordering:
\begin{equation}
{\cal H}_{N}(\{\theta_i \}) = \sum_{i=1}^{N}\lambda_i I_{\theta_i \neq \omega_i},
\label{eq:ham_glass}
\end{equation}
where $I_{A}=1$ if $A$ is true and $I_{A}$ = 0 otherwise, and $(\theta_1, \theta_2, \ldots, \theta_N) \in \text{perm}(\omega_1, \omega_2, \ldots, \omega_N)$. We term the state $\vec{\theta} = \vec{\omega}$ the ``completely correct state," and we say component $k$ of $\vec{\theta}$ is ``incorrectly ordered" if it is not equal to $\omega_k$. The order parameter of our system is $\sum_{i=1}^{N}\langle I_{\theta_i \neq \omega_i} \rangle$,  the average number of incorrectly ordered components of $\vec{\theta}$. We denote this average more succinctly as $\langle j \rangle$. 

We can depict the various microstates of this system as a permutation graph, shown in \reffig{permgraph} for the case $N=15$. In each graph, whenever there is not a line connecting a lattice site $k$ to its vertical complement, the system gains a Boltzmann factor $e^{-\beta\lambda_k}$. The parameter $j$ is defined as the number of diagonal lines, and when $j=0$ we say the system is in the completely correct state.

\begin{figure}[t]
\centering
\includegraphics[width=.95\linewidth]{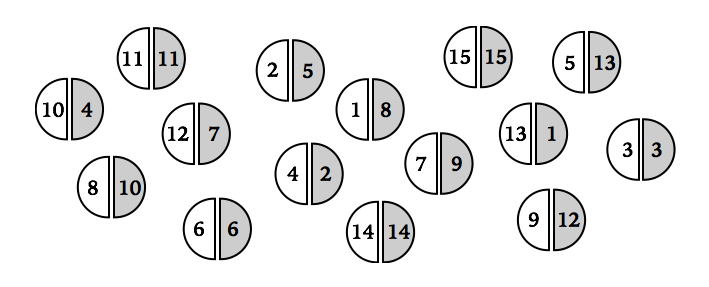}
	\caption{``Matching problem" depiction of a $j=10$ microstate for a $2N = 30$ permutation system. The spatial location of each pair is not important in determining the energy of the state. For this state, the matching pairs are 3, 6, 11, 14, and 15. If we associate a Boltzmann factor $e^{-\beta\lambda_k}$ with each shaded circle $k$ which is not paired with the corresponding unshaded circle $k$, multiply all Boltzmann factors for all pairings within a state, and sum over all possible pairings weighted by the net Boltzmann factor for each collection of pairings, we obtain \refew{partfunc1}.} 
	\label{fig:permsys}
\end{figure}

Beyond a straightforward permutation interpretation of this model, there is an alternative (but formally equivalent) system which is defined by the Hamiltonian \refew{ham_glass}. Consider a collection of $2N$ subunits which only exist in $N$ labeled pairs where each pair consists of a black subunit and a white subunit, e.g., $(B_1, W_1), (B_2, W_2), \ldots (B_{2N}, W_{2N})$. The various microstates of the system (an example of which is shown in \reffig{permsys}) are defined as the various ways the pairings could be arranged while ensuring that each pair has one black and one white subunit. If we associate an energy cost $\lambda_k$ with any pairing where $B_k$ is not paired with its associated $W_k$, then the statistical physics of the system would be identical to the statistical physics of the permutation model governed by \refew{ham_glass}. This amounts to the statistical physics of the ``matching hat" problem \cite{blitzstein2014introduction}. 

In \cite{williams2017statistical}, we found that the statistical physics of such simple systems was quite interesting because even though their properties were governed by the Hamiltonian \refew{ham_glass} in which distinct lattice sites did not interact, we could still have qualitative thermodynamically-driven changes in the properties of our system due to the nature of the state space. More quantitatively, for the system defined by the non-interacting Hamiltonian \refew{ham_glass} with the global energy cost $\lambda_i = \lambda_0$ for all $i$, we can show (Appendix \ref{app2}) that the correlation between the incorrectness of two sites $i$ and $k$ (with $i \neq k$) is  
\begin{equation}
\sigma_{ik}^2  \simeq \frac{1}{N-1}\left( \frac{e^{\beta\lambda_0}}{N}\right)^2,
\label{eq:corr}
\end{equation}
for $\beta\lambda_0 < \ln N$. Thus, as the temperature of our system decreases, the likelihood that component $i$ is incorrectly ordered given that $k$ is incorrectly ordered increases. Therefore, the sites are correlated in spite of the non-interacting Hamiltonian. For systems with correlated degrees of freedom, introducing disorder results in qualitative changes in the system's thermal properties. So we can naturally wonder how disorder would affect a model in which the correlation arises at the level of the state space rather than the Hamiltonian.  

We explore these ideas in a simple model of a permutation glass. We define a permutation glass as a statistical physics system with a state space consisting of various permutations of a list and with a Hamiltonian defined by a quenched distribution of parameter values. In a previous model, we set $\lambda_i = \lambda_0$ for analytic simplicity, but now we will maintain our distribution of $\lambda_i$ values. 

By \cite{williams2017statistical}, the partition function for the system with Hamiltonian \refew{ham_glass} is 
\begin{equation}
Z_{N}(\{\beta \lambda_i\}) =  \int^{\infty}_{0} ds\, e^{-s} \prod^{N}_{\ell=1}\Big[ 1+ (s-1)e^{-\beta\lambda_{\ell}}\Big].
\label{eq:partfunc1}
\end{equation}
Applying Laplace's method to \refew{partfunc1}, we can define the approximate free energy ${\cal F}(s_0)$ (modulo a thermodynamically irrelevant factor) according to 
\begin{equation}
Z_{N}(\{\beta \lambda_i\})  \simeq \exp\left[ - \beta {\cal F}(s_0) \right]
\label{eq:part_free_en}
\end{equation}
where 
\begin{equation}
\beta {\cal F}(s_0) =  s_0 - \sum^{N}_{\ell=1}\ln\Big( 1+ (s_0-1)e^{-\beta\lambda_{\ell}}\Big),
\label{eq:free_en1_perm}
\end{equation}
and where $s_0 = s_0(\beta\lambda_1, \cdots, \beta\lambda_N)$ is defined by the critical point condition 
\begin{equation}
\sum_{\ell=1}^{N} \frac{1}{e^{\beta\lambda_{\ell}} + s_{0}-1} = 1. 
\label{eq:lami}
\end{equation}
We note that the second derivative of the argument of \refew{free_en1_perm} yields
\begin{equation}
\beta {\cal F}''(s_0) = \sum_{\ell=1}^N \frac{e^{-2\beta\lambda_{\ell}}}{(1+(s_0-1)e^{-\beta\lambda_{\ell}})^2},
\label{eq:stable}
\end{equation}
which is always greater than zero. Thus any critical point solution of \refew{lami} is also a stable thermal equilibrium. 

Also, although \refew{free_en1_perm} is in fact an approximation of the true free energy $\beta F = -\ln Z_{N}(\{\beta\lambda_i\})$, henceforth, we will work within our approximation and take \refew{free_en1_perm} to be the free energy from which all thermodynamic quantities are computed. The validity of this approximation is coarsely constrained by parameter regimes over which $s_0$ yields a stable equilibrium for ${\cal F}$, and, by \refew{stable}, this stability is itself only constrained by the physical relevance of the solutions to \refew{lami}. 

We can write \refew{lami} in a more physically transparent form. Noting that the average number of incorrect components $\langle j \rangle$ is the sum of $\langle I_{\theta_i \neq \omega_i} \rangle$ over all components, we have 
\begin{align}
\langle j \rangle & 
= \sum_{k=1}^N \left\langle \frac{\partial (\beta {\cal F})}{\partial (\beta \lambda_{k})} \right \rangle 
 = \sum_{k=1}^{N} \frac{(s_0-1)e^{-\beta\lambda_k}}{1 + (s_0-1)e^{-\beta\lambda_k}}.
\end{align}
Thus, we find

$\langle j \rangle = s_0 -1$, and so \refew{lami} becomes
\begin{equation}
\sum_{\ell = 1}^N \frac{1}{e^{\beta \lambda_{\ell}} + \langle j \rangle } = 1,
\label{eq:lami2}
\end{equation}
where $\langle j \rangle$ is the order parameter of our permutation system. Therefore, \refew{lami2} defines the equilibrium of our system given the set of energy costs $\{\lambda_k\}$. 

\section{TRANSITION TO THE CORRECT MACROSTATE \label{sec:three}}

In order to find the equilibrium behavior  governed by \refew{lami2}, it is useful to introduce a specific distribution of $\lambda_{\ell}$ values and convert \refew{lami2} into an integral. For a sum over a general function $f(\lambda_{\ell})$, where the $\lambda_{\ell}$ are drawn from a normalized distribution $\rho_0(\lambda)$, we can write
\begin{align}
\frac{1}{N} \sum_{\ell=1}^N f(\lambda_{\ell}) &   =  \int^{\infty}_{-\infty} d\lambda f(\lambda) \left[ \frac{1}{N} \sum_{\ell =1}^{N} \delta(\lambda- \lambda_{\ell}) \right]  \mm
&\equiv \int^{\infty}_{-\infty} d\lambda\, f(\lambda)  \rho_0(\lambda). 
\label{eq:distr1}
\end{align}
Thus, with each $\lambda_j$ being drawn from the distribution $\rho_{0}(\lambda)$, \refew{lami2} can be written as 
\begin{equation}
\int^{\infty}_{-\infty} d\lambda\, \frac{\rho_{0}(\lambda)}{e^{\beta\lambda} + \langle j \rangle} = \frac{1}{N}.
\label{eq:lami_3}
\end{equation}

This is not the typical way we start an analysis of glassy systems. Motivated by \cite{sherrington1975solvable}, the typical approach is to use the replica formalism to simplify the partition function and then use a stability analysis to check the validity of the simplification. Fortunately, as shown in Appendix \ref{rsym}, the result \refew{lami_3} is consistent with the condition for the existence of the replica symmetric ansatz of the quenched free energy. More encouragingly, as our distribution $\rho_{0}(\lambda)$ becomes more centered around a single value $\lambda = \lambda_1$, we have $\rho_{ 0}(\lambda) \to \delta(\lambda-\lambda_1)$, which leads to \refew{lami_3} reproducing the non-disordered behavior $\langle j \rangle \simeq N - e^{\beta \lambda_1}$ found in \cite{williams2017statistical}. 

\refew{lami_3}  does not appear any more soluble than \refew{lami2}, but we can use it to derive a general result characterizing one type of temperature-dependent behavior in this system: the thermal transition from $\langle j \rangle \neq 0$ to $\langle j \rangle = 0$. Setting $\langle j \rangle =0$ in \refew{lami_3} for some $\beta_{c}$, we have
\begin{equation}
\int^{\infty}_{-\infty} d\lambda\,\rho_{0}(\lambda)e^{-\beta_c\lambda} = \frac{1}{N}.
\label{eq:lami_3a}
\end{equation}
For a given distribution $\rho_0(\lambda)$, \refew{lami_3a} can be computed and then inverted to find the temperature $k_B T_c = 1/\beta_c$ at which the permutation glass achieves the $\langle j \rangle = 0$ state. But even without detailed knowledge of the distribution, we can use Jensen's inequality \cite{chandler1987introduction} to find an upper limit on this temperature. Given that $e^{x}$ is convex, and defining $\overline{f(\lambda)} \equiv \int d\lambda \rho_{0}(\lambda) f(\lambda)$ we have $\overline{ e^{- \beta_c \lambda}}  \geq e^{- \beta_{c} \overline{\lambda}}$, and thus by \refew{lami_3a} we find
\begin{equation}
k_{B}T_{c} \leq \frac{\overline{\lambda}}{\ln N}.
\label{eq:tem_limit}
\end{equation}
\refew{tem_limit} states that the temperature at which the permutation glass achieves the completely correct $\langle j \rangle = 0$ state is always less than the corresponding temperature predicted from the permutation system in which all interaction terms have the value $\lambda_i = \overline{\lambda}$. In essence, incorporating disorder into the interaction terms leads to a reduced tolerance for thermal disorder in achieving the $\langle j \rangle= 0$ state. Moreover, \refew{tem_limit} indicates that the $\langle j \rangle =0$ state is achievable only if the mean of the $\lambda$ distribution is positive. 

We can derive an approximate expression for this transition temperature in the limit of small disorder. By the fact that the Fourier transform of $\rho_0(\lambda)$ is the exponential of the cumulant generating function \cite{kardar2007statistical}, we find that \refew{lami_3a} implies
\begin{equation}
\sum_{n=1}^{\infty}\frac{(-\beta_c)^n}{n!} \overline{\lambda^n}_c + \ln N = 0
\label{eq:cum_ex}
\end{equation}
where $\overline{\lambda^n}_c$ is the $n$th cumulant of the distribution $\rho_0(\lambda)$. \refew{cum_ex} does not allow us to exactly solve for $\beta_c$ in terms of the cumulants, but it does allow us to solve for $\beta_c$ perturbatively assuming the series is dominated by the first and second cumulant. Noting the first cumulant is the mean $\overline{\lambda}$, the second cumulant is the variance $\sigma_{\lambda}^2$, and assuming $(\beta_c \sigma_{\lambda})^2 \gg \beta_c^k \langle \lambda^k \rangle_c$ for $k >2$, we can approximately solve \refew{cum_ex} to obtain 
\begin{equation}
\beta_c = \frac{\overline{\lambda}}{\sigma_{\lambda}^2} \left( 1 - \sqrt{1 - \frac{2\sigma_{\lambda}^2}{\overline{\lambda}^2} \ln N}\,\right) +\cdots,
\label{eq:bc}
\end{equation}
where we dropped the extraneous solution which yields $\beta_c \to \infty$ as $\sigma_{\lambda} \to 0$.  \refew{bc} is a general result giving the temperature at which $\langle j \rangle = 0$ transitions to $\langle j \rangle \neq 0$ (or vice-versa) for any distribution $\rho_0(\lambda)$, contingent on the assumption that the cumulants of order higher order than $2$ are subdominant. In spite of its limited validity, this result affords us some intuition into how small amounts of disorder affect the transition temperature of our system. If we take our distribution of energy costs to be highly peaked at $\overline{\lambda}$ with a small width $\sigma_{\lambda} \sqrt{2 \ln N} \ll \overline{\lambda} $, we can expand \refew{bc} to find 
 \begin{equation}
\frac{T_c(\sigma_{\lambda})}{T_c(0)} =  1 - \frac{\sigma_{\lambda}^2}{2\overline{\lambda}^2} \ln N + {\cal O}\left(\sigma_{\lambda}^4/\overline{\lambda}^4\right), 
\label{eq:tc}
\end{equation}
where $T_c(0) \equiv T_c(\sigma_{\lambda} = 0) = \overline{\lambda}/\ln N$ is the transition temperature for the non-disordered system. Consistent with \refew{tem_limit},  \refew{tc} shows that the effect of making our permutation system slightly glassy (i.e., imbuing it with nonzero $\sigma_{\lambda}$) is to lower the temperature at which the system transitions from $\langle j \rangle \neq 0$ to $\langle j \rangle =0$.  

The qualitative explanation for this result is straightforward. Introducing disorder at the level of interactions effectively increases the entropy of our system, and the system then compensates for this additional entropy by making the thermal disorder limit for achieving the $\langle j \rangle = 0$ state more stringent. In a sense, because of the interaction disorder, the free energy equilibrium of the system becomes less tolerant of thermal disorder. Thus, the transition temperature, a proxy for limiting thermal disorder, is reduced. A heuristic derivation employing this intuition and reproducing an order of magnitude estimate of \refew{tc} is provided in Appendix \ref{app:heuristic}. 

To generalize this result, we cannot make direct use of the expansion \refew{cum_ex}: Since \refew{bc} and \refew{tc} do not apply when higher-order cumulants cannot be neglected, a perturbative analysis is not generally useful. Therefore when higher order cumulants are relevant, we have to calculate \refew{lami_3} analytically or numerically and then determine how the value and existence of $\beta_c$ depend on the properties of the chosen $\rho_0(\lambda)$. In the next section, we discuss how such properties affect $\beta_c$ by calculating the transition temperature for different energy-cost distributions. 

 \subsection{Example distributions \label{sec:example} }
The result \refew{tc} predicts the value of the transition temperature presuming the width of the energy-cost distribution is small. To find the transition temperature more generally we would need to evaluate \refew{lami_3a} exactly.  We perform this calculation by considering example distributions of $\rho_{0}(\lambda)$: a Gaussian distribution, a uniform distribution, and a symmetric Bernoulli distribution.

In analogy to \refew{bc}, we compute the transition temperature, and, additionally, the conditions for the existence of the transition temperature for each of these distributions. In a later section, we will show that in spite of the diversity of these conditions, they can all be subsumed into a single inequality which places an upper limit on the probability that our energy cost for an incorrect component is less than zero (i.e., the probability that the energy cost is actually an energy benefit). 

\subsubsection{Gaussian distribution}
We consider a Gaussian distribution. Given mean $\lambda_0$ and variance $\sigma_{0}^2$, we have the energy-costs density
\begin{equation}
\rho_{0}(\lambda) = \frac{1}{\sqrt{2\pi \sigma_{0}^2}} e^{-(\lambda- \lambda_0)^2 /2\sigma_{0}^2}. 
\label{eq:distr3}
\end{equation}
With this distribution, we would like to use \refew{lami_3} to find a closed-form analytic expression for $\langle j \rangle$, but, due to the insolubility of the resulting integral, we will instead use \refew{cum_ex} to find a value for $\beta_c$. Because \refew{distr3} is Gaussian, the cumulants of order higher than $2$ are zero, and \refew{cum_ex} reduces to 
\begin{equation}
- \beta_c \lambda_0 + \frac{1}{2}\beta_c^2 \sigma_{\lambda}^2 + \ln N = 0.
\label{eq:gauss_constraint}
\end{equation}
Therefore we find our $\beta_c$ is exactly identical to \refew{bc} without the additional higher order terms: 
\begin{equation}
\beta_c = \frac{\lambda_0}{\sigma_{0}^2} \left( 1 - \sqrt{1 - \frac{2\sigma_{0}^2}{\lambda_{0}^2} \ln N}\,\right).
\label{eq:bc_gauss}
\end{equation}
A corollary of \refew{bc_gauss} is that $\beta_c$ exists and the system is able to achieve the $\langle j \rangle = 0$ state only if the mean and variance of the Gaussian satisfy
\begin{equation}
\frac{\lambda_0 }{\sigma_{0}} \geq \sqrt{2 \ln N}.
\label{eq:gauss_ineq}
\end{equation}
\refew{gauss_ineq} indicates that as $N\to \infty$ and the number of incorrect microstates in the system increases, the mean of the Gaussian distribution of energy costs must increase with $N$--although sub-logarithmically so--in order for the $\langle j \rangle =0$ state to be achievable. 

\subsubsection{Uniform distribution} 
We consider a uniform distribution with a finite domain. The distribution of $\lambda$ values is defined as 
\begin{equation}
\rho_{0}(\lambda) = \begin{dcases}\frac{1}{2\sqrt{3}\sigma_{0}} & \text{ for $\lambda_0 - \sigma_{0} \sqrt{3} \leq \lambda \leq \lambda_0 +\sigma_{0} \sqrt{3}$ } \\ 
0 & \text{ otherwise}. \end{dcases} 
\label{eq:distr2}
\end{equation}
\refew{distr2} defines a system in which each $\lambda_k$ in \refew{ham_glass} has a constant probability $\Delta \lambda/2\sqrt{3}\sigma_0$ to be found within an energy width $\Delta \lambda$ as long as this width is within the domain $[\lambda_0 - \sigma_0 \sqrt{3}, \lambda_0 + \sigma_0 \sqrt{3}]$.
The form of \refew{distr2} was chosen so that the mean is $\lambda_0$ and the variance is $\sigma_{0}^2$.  For this distribution, we cannot compute $\langle j \rangle$ exactly given \refew{lami_3}, but we can establish an implicit condition on the existence of $\langle  j \rangle = 0$.  Computing \refew{lami_3a} given \refew{distr2}, and, taking the logarithm of the result, we find the condition 
\begin{equation}
-\beta_c \lambda_0+\ln \left[\frac{\sinh(\beta_c \sigma_{0}\sqrt{3})}{\beta_c \sigma_{0}\sqrt{3}}\right] + \ln N = 0  .
\label{eq:flat_ineq}
\end{equation}
We note that as $\sigma_{0} \to 0$ in \refew{flat_ineq}, $\beta_c \to \ln N/\lambda_0$, and thus this result is consistent with the zero-disorder limit. Moreover, if we were to expand \refew{flat_ineq} in the limit $\beta_c \sigma_0 \ll 1$, we would obtain a quadratic equation the solution of which matches \refew{bc}. Considering the large-disorder limit $\beta_c \sigma_0 \gg1$, we find that \refew{flat_ineq} has the solution 
\begin{equation}
\beta_c \simeq \frac{1}{\lambda_0 - \sigma_0 \sqrt{3}} W_{0} \left(\frac{N}{2\sigma_{0}\sqrt{3}}(\lambda_0 - \sigma_0 \sqrt{3}) \right),
\label{eq:flat_ineq2}
\end{equation}
where $W_0(x)$ is the principal branch of the Lambert-$W$ function \cite{weisstein2002lambert}. Since the sign of $W_0$ matches the sign of its argument, \refew{flat_ineq2} is always positive for valid ranges of the distribution parameters. Thus the parameters are only constrained by the existence of a real $W_0$, which is in turn constrained by the condition that its argument is greater than or equal to $-e^{-1}$. We therefore find that for \refew{flat_ineq2} to exist (and hence for the system to be able to achieve the $\langle j \rangle=0$ state), the mean and variance must satisfy
\begin{equation}
\frac{\lambda_0}{\sigma_0} \geq \sqrt{3}\left(1- \frac{2}{N e} \right).
\label{eq:flat_ineq3}
\end{equation}
We note that \refew{flat_ineq3}, in contrast to \refew{gauss_ineq}, becomes independent of $N$ in the $N\gg1$ limit. Namely, as $N\to\infty$, the mean of the uniform distribution just needs to exceed a fixed multiple of the variance in order for the $\langle j \rangle = 0$ state to be achievable.

\subsubsection{Symmetric Bernoulli distribution}
We consider a symmetric Bernoulli distribution. The energy costs are distributed according to
\begin{equation}
\rho_{0}(\lambda) = q \delta(\lambda - \lambda_{+}) + (1-q) \delta(\lambda+ \lambda_{+}),
\label{eq:distr1a}
\end{equation}
where $q$ is a dimensionless number satisfying $0 \leq q \leq 1$, and we take $\lambda_{+} >0$. Conceptually, \refew{distr1a} defines a system in which each $\lambda_k$ in \refew{ham_glass} has a probability $q$ of being $\lambda_{+}$ and a probability $1-q$ of being $-\lambda_{+}$. It is possible to solve \refew{lami_3} for $\langle j \rangle$ given the distribution \refew{distr1a} (Appendix \ref{app:bernoulli}), but here we are more concerned with the conditions which allow for the existence of $\langle j \rangle = 0$. 

For \refew{distr1a}, the conditions for the existence of a $\beta_c$ satisfying \refew{bc} are 
\begin{equation}
q e^{-\beta_c\lambda_{+}}+(1-q)e^{\beta_c\lambda_{+}} = \frac{1}{N}. 
\label{eq:distr1ab}
\end{equation}
As a check, we note that taking $q \to 1$ in \refew{distr1ab} yields the solution $\beta_c = \ln N/ {\lambda_{+}}$ as expected. Also, taking the logarithm of both sides of \refew{distr1ab} and expanding the right hand side to second order in $\beta_c$ yields a quadratic equation which reproduces \refew{bc} upon solution.

\refew{distr1ab} can be solved exactly for $\beta_c$. Doing so (and dropping the solution which does not yield a finite $\beta_c$ in the $q \to 1$ limit) yields 
\begin{equation}
\beta_c \lambda_{+} = \ln \left[ \frac{1}{2N(1-q)} \left( 1 - \sqrt{1- 4N^2 q (1-q)}\right) \right].
\label{eq:bc0}
\end{equation}
It is possible to show that the argument of the logarithm in \refew{bc0} is always greater than 1 provided $N>1$. Thus, the only constraint on the existence of a real and positive $\beta_c$ is the sign of the argument in the square root. Mandating the argument of the square root is positive semidefinite, we find the condition
\begin{equation}
q \geq  \frac{1}{2} \left( 1 + \sqrt{1- \frac{1}{N^2}} \right),
\label{eq:refq}
\end{equation}
where we dropped the inequality which allowed for an extraneous $q=0$ solution. In the limit $N\gg1$, \refew{refq} tells us that the distribution \refew{distr1a} only yields the completely correct $\langle j \rangle = 0$ equilibrium if the probability of getting $\lambda = \lambda_{+}$ is very close to 1. 

The minimal $q$ predicted by \refew{refq} can be understood from the form of \refew{distr1a}. The two $\lambda$ values permitted by \refew{distr1a} are symmetric about $\lambda = 0$, but because the existence of the equilibrium $\langle j \rangle= 0$ depends only on the existence of an energy cost (rather than an energy benefit) of deviating from the correctly ordered microstate, only the positive $\lambda$ value ensures the possibility of the $\langle j \rangle =0$ equilibrium. As $N$ increases, the possible number of  incorrectly-ordered states in the system increases, and thus to ensure that all components are on average correctly ordered (i.e., that $\langle j \rangle  = 0$ is satisfied), there needs to be a greater probability of having an energy cost and a corresponding lower probability of having an energy benefit. Thus we find $q$ must approach $1$ as $N \to \infty$. 

To compare \refew{refq} with the results for our other distributions, we rewrite it in terms of the mean $\lambda_0$ and variance $\sigma_{0}^2$. From \refew{distr1a}, we find 
\begin{equation}
\lambda_0 = \lambda_{+}(2q-1), \quad \sigma_{0}^2 = \lambda_{+}^2 4q(1-q).
\label{eq:avg_sig}
\end{equation}
Using \refew{avg_sig} to translate the inequality \refew{refq} into a constraint on $\lambda_0$ and $\sigma_0$, we find that $\beta_c$ in \refew{bc0} only exists if 
\begin{equation}
\frac{\lambda_0}{\sigma_{0}} \geq \sqrt{N^2 - 1}. 
\label{eq:dirac_ineq}
\end{equation}
In other words, \refew{dirac_ineq} establishes the condition the distribution \refew{distr1a} must satisfy in order for the system to admit a $\langle j \rangle =0$ equilibrium. Comparing \refew{dirac_ineq}, \refew{gauss_ineq},  and \refew{flat_ineq3} we note that \refew{dirac_ineq} establishes the most stringent constraint for the existence of this equilibrium: As $N\to \infty$, the mean energy costs must increase linearly with $N$ in order for the $\langle j \rangle =0$ state to be achievable.

\subsection{Comparison of transition temperatures}

\begin{figure}[t]
\includegraphics[width=.85\linewidth]{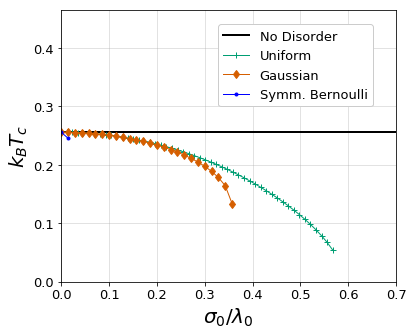}
	\caption{Plots of transition temperature vs mean-normalized standard deviation for $N=50$ and $\lambda_0 =1.0$. The horizontal axis is a proxy for disorder in the energy-cost distribution, and the vertical axis gives the corresponding thermal disorder the system can tolerate while still allowing $\langle j \rangle=0$ to be accessible. The $k_BT_c$ line for no disorder is $\lambda_0/\ln N$. The $k_BT_c$ curves for the Gaussian, uniform, and symmetric Bernoulli distribution are found from \refew{bc_gauss}, \refew{flat_ineq}, and \refew{bc0}, respectively. Each curve is only plotted for the domain of $\sigma_0/\lambda_0$ which yields real values for $k_BT_c$, and thus the relative end points of the curves allow us to compare which distributions are most tolerant of disorder in the energy-cost distribution. 
	} 
	\label{fig:Tc_disorder}
\end{figure}

In \reffig{Tc_disorder} we plot the derived transition temperatures \refew{bc_gauss}, \refew{flat_ineq}, and  \refew{bc0} (with $\lambda_{+}$ and $q$ computed from \refew{avg_sig}) as functions of $\sigma_0/\lambda_0$. 
We see that the symmetric Bernoulli distribution curve ends at $\sigma_0/\lambda_0 \approx 0.02$  and thus admits  the smallest amount of energy-cost disorder before the $\langle j \rangle =0$ state is unachievable. Conversely, the uniform distribution ends at $\sigma_0/\lambda_0 \approx 0.56$ and thus admits the largest amount of energy-cost disorder. 

Consistent with \refew{tem_limit} and the intuition underlying \refew{tc}, we find that each distribution predicts a transition temperature satisfying
\begin{equation}
k_BT_c \leq \frac{\lambda_0}{\ln N}
\label{eq:trans_ineq}
\end{equation}
and thus predicts a lower transition temperature than the corresponding non-disordered prediction. In the same way that the transition temperature results  \refew{bc_gauss}, \refew{bc0}, and \refew{flat_ineq} must be consistent with \refew{trans_ineq}, in a future section, we will show how each of the previous distribution-specific constraints on  $\lambda_0/\sigma_0$ can be unified into a distribution-independent result expressed in terms of $P_{\lambda<0}$, the probability that an incorrectly-ordered component is energetically favored. But first, in the next section, we compare these analytic results to results from simulations. 
\section{SIMULATION COMPARISON\label{sec:sim}}

\begin{figure*}[t]
  \centering
\begin{subfigure}[t]{0.325\textwidth}
  \centering
  \includegraphics[width=\linewidth]{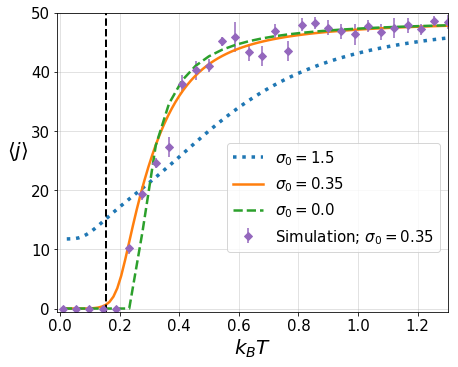}
	\caption{}
	\label{fig:avg_j_gauss}
\end{subfigure}
\begin{subfigure}[t]{0.325\textwidth}
\centering
\includegraphics[width=\linewidth]{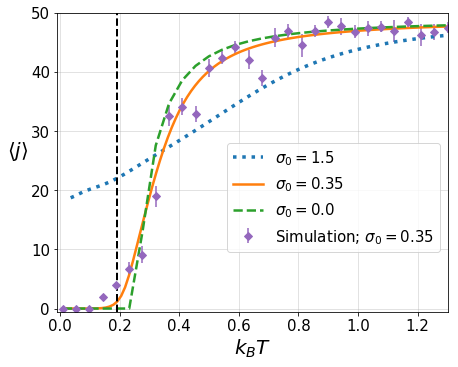}
	\caption{}
	\label{fig:avg_j_flat}
\end{subfigure}
\begin{subfigure}[t]{0.325\textwidth}
\centering
\includegraphics[width=\linewidth]{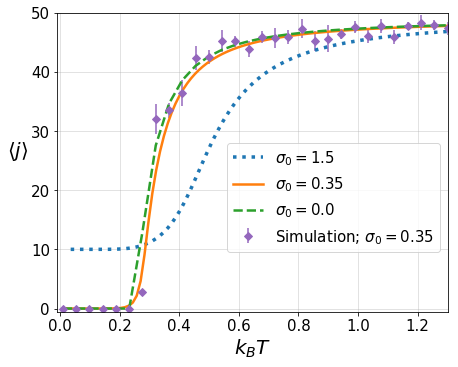}
	\caption{}
	\label{fig:avg_j_dirac}
\end{subfigure}
	\caption{
	Theory and simulation comparison for $N=50$. (a) Gaussian distribution. (b) Uniform distribution. (c) Symmetric Bernoulli distribution. All distributions were defined with $\lambda_0 = 1.0$. As the theoretical (i.e., non-simulated) results, in each figure, we have dashed, dotted, and solid line curves which were all computed from \refew{ord_exact}. The dashed curve corresponds to the zero-disorder ($\sigma_0=0.0$) solution \refew{jsimp}. The dotted curve corresponds to the high-disorder ($\sigma_0/\lambda_0 = 1.5$) solution. The solid curve defines our intermediate-disorder ($\sigma_0 =0.35$) solution. The vertical dashed lines in (a) and (b) are the transition temperatures computed from  \refew{bc_gauss} and \refew{flat_ineq2}, respectively, at $(\lambda_0, \sigma_0) = (1. 0, 0.35)$; there is no transition temperature for \reffig{avg_j_dirac} because $\sigma_0/\lambda_0 = 0.35$ violates \refew{dirac_ineq}. The points in each plot are the simulated results for the $\sigma_0 = 0.35$ solution. We note that for (a), (b), and (c), \refew{ord_exact} matches the simulated results, and for (a) and (b), the temperatures computed from \refew{bc_gauss} and \refew{flat_ineq2} correctly match the temperature value at which the simulation transitions from $\langle j \rangle=0$ to $\langle j \rangle >0$. Code for generating figure is linked to in \textit{Supplementary Code} (\refsec{supp}).
}
	\label{fig:simulation2}
\end{figure*}

We seek to affirm the theoretical transition temperatures of the previous section by comparing them to simulation results. Doing so requires us to simulate how $\langle j \rangle$ varies as a function of temperature when the $\{\lambda_i\}$ are drawn from the Gaussian, uniform, and symmetric Bernoulli distributions. 

First, we derive a more general theoretical prediction to which we will compare the simulations. For the statistical physics of the non-disordered model where the Hamiltonian is ${\cal H} = \lambda_0 \sum_{i} I_{\theta_i \neq \omega_i}$, we know the order parameter has the simple form 
\begin{equation}
\langle j \rangle \simeq N - e^{\beta\lambda_0} \qquad \text{(non-disordered result)}.
\label{eq:jsimp}
\end{equation}
We would like to find analogous results for our permutation glasses defined by various distributions of energy costs. This would amount to computing \refew{lami_3a} for a given distribution and inverting it to find $\langle j \rangle$ as a function of temperature and the parameters of the distribution. This procedure can be implemented exactly for the symmetric Bernoulli distribution, but there seems to be no analytic solution for the Gaussian or the uniform distribution. So, we will instead use a more general expression for $\langle j \rangle$ which allows us to reduce all the distribution-dependent order parameters to a common form. 

Given the Hamiltonian \refew{ham_glass} and that $\langle j \rangle$ is the sum of $\langle I_{\theta_i \neq \omega_i} \rangle$ across all components, we have 
\begin{equation}
\qquad \langle j \rangle = -\sum_{i=1}^{N} \frac{\partial}{\partial(\beta \lambda_i)} \ln Z_{N}(\{\beta\lambda_i\}).
\end{equation}
Then, using \refew{partfunc1},  yields the general result
\begin{align}
\langle j \rangle &= N - \frac{1}{Z_{N}(\{\beta \lambda_i\}) }\sum_{k=1}^NZ_{N-1}(\{\beta \lambda_i\}_{i \neq k}),
\label{eq:ord_exact}
\end{align}
where $Z_{N-1}(\{\beta \lambda_i\}_{i \neq k})$ is defined by \refew{partfunc1} with the product taken over the $N-1$ elements of $\{\lambda_i\}$ not including $\lambda_k$. The utility of \refew{ord_exact} is that it gives us the exact temperature dependence of the order parameter contingent on a particular distribution of energy costs. The caveat is that, rather than being a function of distribution parameters like means and variances, \refew{ord_exact} requires us to draw the explicit set of $\{\lambda_i\}$ from the given distribution.

With \refew{ord_exact}, we have our theoretical prediction and can now discuss the simulation. The simulation was set up as follows: First,  the vector $\vec{\omega} = (1, 2, \ldots, N)$ was defined to be the completely correct permutation. This was the initial state in the simulation. Single-step state transitions were enacted by exchanging two randomly chosen elements of the current vector contingent on the Metropolis acceptance criterion, i.e., that $e^{-(E_f- E_i)/T}< u$ where $E_f$ and $E_i$ were the final and initial state energies, respectively, $T$ was the temperature, and $u$ was a number drawn uniformly from $[0, 1)$. The initial and final state energies were computed from \refew{ham_glass} where the $\{\lambda_i\}$ were drawn from the given distribution defined by a mean $\lambda_0$ and variance $\sigma_0^2$. The simulation was run for $5\times10^4$ steps of which the last $10^3$ steps were used to define the ensemble of states. From this ensemble of states, we computed $j$ (the number of elements in the state which did not match the corresponding element in $\vec{\omega}$) for each state and then averaged this value of $j$ across all states in the ensemble to find the simulation prediction of $\langle j \rangle$ at a specific temperature. We chose $30$ temperature values between $0.1$ and $1.3$. Finally, for a given distribution, the drawn set of $\{\lambda_i\}$ was used in \refew{ord_exact} to obtain the  corresponding theoretical prediction.

In Figures 4a, 4b, and 4c, we show simulation results for the parameter values $(N, \lambda_0, \sigma_0) = (50, 1.0, 0.35)$;  respectively, these figures correspond to the Gaussian, uniform, and symmetric Bernoulli distribution of energy costs.  As theory comparisons, for each distribution, we also plotted \refew{ord_exact} for the same parameter values as in the simulation. We note that in all three cases, the theory curves well match the simulated results. As zero-disorder and high-disorder comparisons, we included theory curves of the order parameter for the standard deviation values $\sigma_0 = 0.0$ and $1.5$ with $N$ and $\lambda_0$ the same in all cases. From the differences in the curves among the plots, we see that at high disorder, the temperature behavior of \refew{ord_exact} is greatly dependent on the distribution from which the $\{\lambda_i\}$ are drawn.

Finally, for \reffig{avg_j_gauss} and \reffig{avg_j_flat}, we computed the transition temperatures obtained from \refew{bc_gauss} and \refew{flat_ineq2}, respectively, and displayed the predictions as vertical dashed lines. Consistent with the simulation results, these lines correspond to the temperature values at which $\langle j \rangle$ transitions from zero to non-zero values or vice versa. Moreover, we note that, consistent with \reffig{Tc_disorder}, a disorder of $\sigma_0/\lambda_0 = 0.35$ allows the order parameter for the Gaussian and uniform distributions to reach $\langle j \rangle =0$ at sufficiently low temperatures, but, at this level of disorder, the order parameter for the symmetric Bernoulli distribution remains non-zero over its entire temperature range because its $k_BT_c$ does not exist. 

The similarity between the theoretical and the simulation results is reassuring, but what still remains is the task of finding a unified interpretation for the constraints \refew{gauss_ineq}, \refew{flat_ineq3}, and \refew{dirac_ineq}. We turn to developing such an interpretation in the following section.

\section{UNDERSTANDING PARAMETER CONSTRAINTS \label{sec:understand}}
What is strange about the parameter conditions given by \refew{gauss_ineq}, \refew{flat_ineq3}, and \refew{dirac_ineq} is their variety.  Although each represents the conditions the mean and variance of the respective distribution must satisfy in order for the $\langle j \rangle =0$ state to be an equilibrium, they all have quite different scaling behaviors as functions of $N$. Perhaps most interestingly, the condition \refew{flat_ineq3} becomes independent of $N$ in the $N \gg 1$ limit, thus suggesting that at large $N$ the amount of interaction disorder a system with a uniform distribution of energy costs can tolerate is independent of the number of microstates available to it. 

However, underlying this variety in parameter conditions is a unity of the situations giving rise to them. Specifically, the conditions \refew{gauss_ineq}, \refew{flat_ineq3}, and \refew{dirac_ineq} are the translations into distribution-parameter language of something which bears a common form when written as a probability. We can understand this by determining how the derived conditions place upper limits on the probability of obtaining an energy benefit, i.e., of drawing  $\lambda<0$ from the distribution. 

We begin with our previous constraint which must be satisfied in order for $\beta_c$ to exist:
\begin{equation}
\int^{\infty}_{-\infty} d\lambda\,\rho_{0}(\lambda)e^{-\beta_c\lambda} = \frac{1}{N}.
\label{eq:main_identity}
\end{equation}
Next, we define
\begin{equation}
P_{\lambda<0} \equiv \int^{0}_{-\infty}d\lambda\, \rho_{0}(\lambda),
\end{equation}
which represents the probability that a $\lambda_k$ in \refew{ham_glass} is less than zero (i.e., yields an energy benefit for an incorrectly ordered component rather than an energy cost). With the fact that $f(x) < e^{x} f(x)$ for $0<x$ and from \refew{main_identity}, we can infer that in order for $\beta_c$ to exist (and, in turn, for the completely correct equilibrium $\langle j \rangle =0$ to be a physical state) we must have
\begin{equation}
P_{\lambda<0} < \frac{1}{N}.
\label{eq:prob_lam}
\end{equation}
Thus as $N\to \infty$, the probability of each lattice site having $\lambda <0$ must go to zero. Physically, we can interpret this result with the same intuition used to interpret \refew{refq}. As the number of sites $N$ in our system increases, the number of potential incorrectly ordered microstates also increases, and thus to combat the entropic disorder from these microstates and to ensure the existence of the $\langle j \rangle =0$ equilibrium, the system must be ever more likely to have an energy cost (rather than an energy benefit) for incorrectly occupying a single site. Thus as $N$ increases, the system must become less tolerant of $\lambda<0$ values, and $P_{\lambda<0}$ goes to zero. Finally,  the probability limit \refew{prob_lam} is consistent with temperature limit \refew{tem_limit} since both inequalities are derived from the same equilibrium condition. 

\begin{figure}[b]
\begin{flushleft}
\includegraphics[width=.9\linewidth]{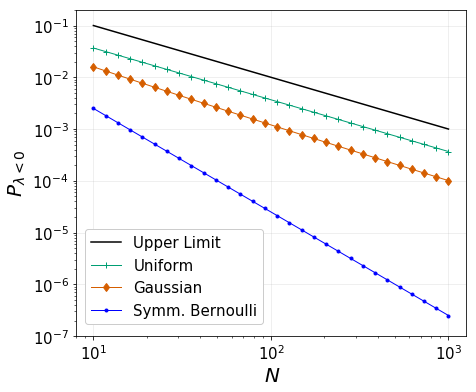}
\end{flushleft}
	\caption{Log-log plot of the critical probabilities \refew{prob_lam},  \refew{probf}, \refew{probg}, and \refew{probd} as functions of $N$. The curves give the critical probability above which the $\langle j \rangle =0$ equilibrium cannot be achieved for the given distribution. The solid curve represents the upper limit on critical probabilities established by \refew{prob_lam}. The closer the probability curve is to this upper limit, the more disorder it can admit before the $\langle j \rangle=0$ equilibrium is unachievable. Consistent with \refew{prob_lam}, each critical probability curve exists below this $1/N$ limit. }
	\label{fig:logprob_plot}
\end{figure}

\refew{prob_lam} is a general result which must be true regardless of the distribution we choose, but what we find is that our previously derived mean-variance conditions are simply representations of \refew{prob_lam} in the language of the parameters which define each specific distribution.  To better understand how our mean-variance conditions \refew{gauss_ineq}, \refew{flat_ineq3}, and \refew{dirac_ineq} are related to \refew{prob_lam}, we interpret them as placing upper limits on how much variance $\sigma_0^2$ the system can tolerate before the $\langle j \rangle =0$ state is no longer an equilibrium.  Given that the $\langle j \rangle =0$ state is only achieved through the positive $\lambda$ domain of the distribution $\rho_0(\lambda)$, the upper limit on $\sigma_0$ must be tantamount to a lower limit on $\int^{\infty}_{0}d\lambda\,\rho_{0}(\lambda)$, or, equivalently an upper limit on $\int_{-\infty}^{0}d\lambda\, \rho_{0}(\lambda)$. Such an upper limit implies that if too much of the distribution is contained within the negative $\lambda$ domain, then the $\langle j \rangle=0$ state is not accessible. Thus interpreting \refew{gauss_ineq}, \refew{flat_ineq3}, and \refew{dirac_ineq} as upper limits on the variances of their respective distributions, we can compute corresponding upper limits on the probability of obtaining a negative value of $\lambda$. For the relevant distributions we find (Appendix \ref{app:prob_distr})
\begin{align}
P^{\text{uniform}}_{\,\lambda<0} &\, \leq \,\frac{1}{Ne} \label{eq:probf} \\
P^{\text{gauss}}_{\,\lambda<0} \,& \lesssim  \,\frac{1}{2 N \sqrt{\pi \ln N}} \label{eq:probg}\\
P^{\text{bernoulli}}_{\,\lambda<0} \,& \lesssim \,\frac{1}{4N^2}\label{eq:probd}
\end{align}
The above expressions represent the maximum probability of having an energy benefit in the system and still being able to achieve the $\langle j \rangle =0$ state at a certain temperature. All of these results are unified by their inverse scaling with $N$ and, as shown in \reffig{logprob_plot}, their consistency with the limit established by \refew{prob_lam}.

The results \refew{probf}, \refew{probg}, and \refew{probd} afford us a new interpretation of the results in \reffig{Tc_disorder}. We previously noted that the uniform distribution admitted the most amount of disorder before the $\langle j \rangle =0$ state was inaccessible and that the symmetric Bernoulli distribution admitted the least amount of disorder. From \reffig{logprob_plot} we see why: The uniform distribution allows the most amount of disorder because it admits the largest probability of energy benefits for incorrectly ordered components. By admitting a larger probability of energetically beneficial incorrect components, the distribution need not be tightly concentrated about the mean and can therefore have a higher standard deviation. Conversely, the symmetric Bernoulli distribution allows the least amount of disorder because it admits the smallest probability of energy benefits for incorrectly ordered components. The Gaussian distribution admits an intermediate value of disorder because its limiting probability exists between the limiting probabilities of the two other distributions. 

Arguably, this explanation simply translates the old question into a new one: Why, conceptually, do the various distributions have the limiting probabilities shown in \reffig{logprob_plot}? Their relative ordering could be understood by considering the $\lambda<0$ tails of each distribution. In order for the $\langle j \rangle =0$ state to be accessible, the distribution needs to be dominated by positive values of $\lambda$. We can roughly understand this by noting that in the non-disordered result \refew{jsimp}, $\langle j \rangle =0$ is not accessible if $\lambda_0<0$. For the uniform distribution \refew{distr2}, it is possible to completely eliminate $\lambda<0$ values by simply increasing the ratio $\lambda_0/\sigma_0$ with $\sigma_0$ finite; thus for the uniform distribution, $\sigma_0$ can be finite and possibly large while the $\langle j \rangle=0$ state is still accessible. However, the long tail of the Gaussian \refew{distr3} implies there will always be $\lambda<0$ values for non-zero $\sigma_0$. This is even more so for the symmetric Bernoulli distribution \refew{distr1a} since its probability density is not defined by an exponential fall off. Thus, in order to limit the $\lambda<0$ values, the Gaussian distribution needs to be less tolerant of large spreads than the uniform distribution, and the symmetric Bernoulli distribution must be even less tolerant than the Gaussian distribution. This relative tolerance of disorder leads to the sequence shown in \reffig{logprob_plot}.

Lastly, noting that the results \refew{probf}, \refew{probg}, and \refew{probd} all scale at least as $\sim 1/N$  with corrections to the power of $N$ contingent on the distribution, we could guess there exists a stronger limit than \refew{prob_lam} which any distribution must satisfy in order for $\beta_c$ to exist. Namely, in order for the $\langle j\rangle =0$ state to be achieved,  we could conjecture that the probability of obtaining an energy benefit must satisfy, in the $N\gg 1$ limit, 
\begin{equation}
P_{\lambda<0} \lesssim \frac{1}{N^{1+\alpha(N)}}, \quad \text{[Conjecture]}
\end{equation}
where $\alpha(N)>0$ is dependent on the properties of the distribution. 

In summary, the variety of results in the conditions placing limits on the distribution parameters is somewhat misleading because what is important is not the parameters themselves but the probabilities (specifically the probability of an energy benefit) they are associated with.  

\begin{figure*}[t]
  \centering
\begin{subfigure}[t]{0.45\textwidth}
\centering
\includegraphics[width=\linewidth, left]{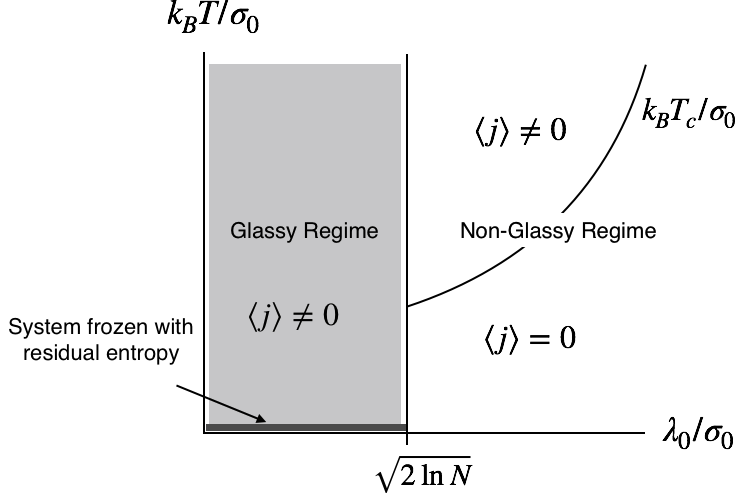}
	\caption{}
	\label{fig:perm_glass_phase}
\end{subfigure}
\hspace{2.0cm}
\begin{subfigure}[t]{0.3\textwidth}
\centering
\includegraphics[width=\linewidth]{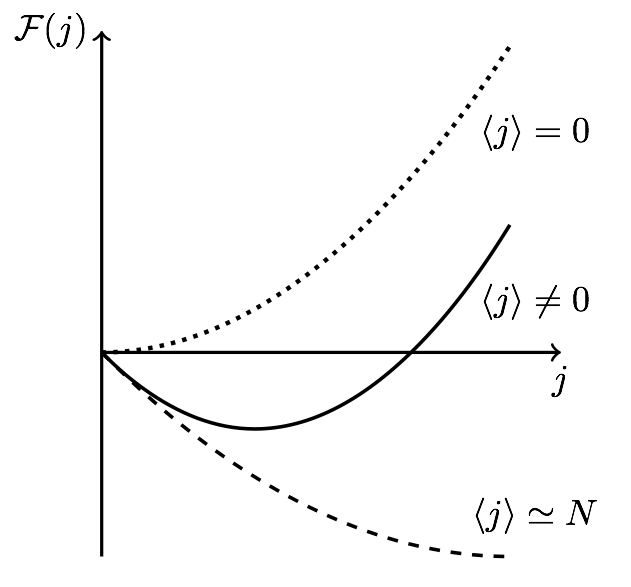}
	\caption{}
	\label{fig:phases}
\end{subfigure}
	\caption{ (a) Glassy and non-glassy regime for a quenched Gaussian distribution of energy costs. Although the order parameter in the non-glassy regime has behavior similar to that in the non-disordered system, the order parameter in the glassy regime is always non-zero and the corresponding system becomes frozen at zero-temperature and exhibits a residual entropy. (b) Schematic of Landau free energy plots of \refew{free_en_perm}. The free energy is shifted so that $\beta {\cal F}(j=0) = 0$. Much like the possible free energies for the non-disordered system where $\rho(\lambda) = \delta(\lambda- \lambda_0)$ (discussed in \cite{williams2017statistical}), we find the disordered system defined by the Hamiltonian \refew{ham_glass} has three possible free energy curves each defined by a single equilibrium $\langle j \rangle$ which falls within $[0, N]$. However, for a system in the glassy regime, the free energy with a global minimum at $\langle j \rangle=0$ cannot be achieved at any temperature even if the average energy cost $\overline{\lambda}$ is positive. Thus the glassy regime is characterized by a non-zero entropy even at zero-temperature.}
	\label{fig:glass_phase_img}
\end{figure*}

\section{GLASSY REGIME \label{sec:four}}

In the previous sections, we considered the conditions defining a permutations glass which allow for the existence of the $\langle j \rangle = 0$ equilibrium. We pursued this analysis in analogy to the non-disordered system where the $\langle j \rangle = 0$ state (found at temperatures below $\lambda_0/\ln N$) defined the only thermal transition in the system. From the discussion in Sec. \ref{sec:understand}, we see that the permutation glass differs from the non-disordered system in that, for the permutation glass, it is possible to have a positive mean energy cost $\overline{\lambda}$ with the system still not transitioning to $\langle j \rangle=0$ at a physical temperature. The $\langle j \rangle=0$ macrostate is significant because it is the only macrostate for which the system has a single microstate and hence an entropy of zero. The number of microstates associated with a general $j$ is given by \cite{williams2017statistical}
\begin{equation}
\Omega_N(j) = \binom{N}{j} d_j, 
\label{eq:degen}
\end{equation}
where $d_j$ is the number of derangements of a list with $j$ elements. We only find $\Omega_N =1$, when $j=0$ and thus if \rfw{main_identity} cannot be satisfied at a physical temperature $k_BT_c$, the system always has a non-zero entropy.

Using \rfw{main_identity}, there are two results which are important in defining a glassy regime for this system. First, we recall the derived inequality 
\begin{equation}
k_BT_c \leq \frac{\overline{\lambda}}{\ln N}.
\label{eq:jensen}
\end{equation}
\rfw{jensen} requires that any temperature at which $\langle j \rangle=0$ is achieved to fall below ${\overline{\lambda}}/{\ln N}$ and, importantly, a necessary condition for such a temperature to exist is for $\overline{\lambda}>0$. For the case without quenched disorder (i.e., $\rho_{0}(\lambda) = \delta(\lambda-\overline{\lambda})$), the inequality in \rfw{jensen} becomes an equality, and the system always assumes the $\langle j \rangle=0$ macrostate when $k_BT$ falls below the stated value.

However, for quenched distributions an additional constraint must be satisfied. For the Gaussian, symmetric Bernoulli, and uniform two-parameter distributions analyzed in Sec \ref{sec:example}, the critical temperature takes on the general schematic form of $k_BT_c(\lambda_0, \sigma_0, N),$ where $\lambda_0$ and $\sigma_0$ are the mean and standard deviation of the distributions. For all of these analyzed distributions, we found that there was a minimum value of $\lambda_0/\sigma_0$ below which $k_BT_c(\lambda_0, \sigma_0, N)$ was no-longer physical. Namely, \rfw{main_identity} has a real solution for $\beta_c$, if and only if 
\begin{equation}
\frac{\lambda_0}{\sigma_0} \geq f(N),
\label{eq:lam_cond}
\end{equation}
for some function $f(N)$ that depends on the properties of the distribution. In other words, even if $\overline{\lambda}>0$ and there are temperatures $k_BT$ that exist below $\overline{\lambda}/\ln N$, none of these temperatures would yield the $\langle j \rangle=0$ equilibrium unless \rfw{lam_cond} is satisfied as well. If \rfw{lam_cond} is violated, then even at zero-temperature we would have $\langle j \rangle\neq0$ and, by \rfw{degen}, the entropy of the system would be non-zero. Therefore, when \rfw{lam_cond} is violated the system exhibits a zero-temperature residual entropy typical of glassy systems \cite{nemilov2009zero} and we can take a violation of \rfw{lam_cond} together with a positive $\overline{\lambda}$ to be definitive of the "glassy regime" of the system. 

These ideas become clearer with a concrete example and a figure. For a quenched distribution of energy-costs drawn from a Gaussian distribution, we found the critical temperature.
\begin{equation}
\frac{k_BT_c}{\sigma_0} = \left(\frac{\lambda_0}{\sigma_0}- \sqrt{\frac{\lambda_0^2}{\sigma_0^2}- 2 \ln N}\,\,\right)^{-1}.
\label{eq:phys_temp}
\end{equation}
The phase diagram associated with this result is shown in \reffig{perm_glass_phase}. The figure depicts the fact that, for $\lambda_0/\sigma_0< \sqrt{2\ln N}$, \rfw{phys_temp} becomes imaginary and the system enters the glassy regime and that although both the non-glassy and glassy regime have $\langle j \rangle\neq 0$ macrostates, unlike the non-glassy regime, the glassy regime never achieves $\langle j \rangle=0$ at a physical temperature. 

We have already derived a more general condition for establishing the existence of the glassy regime. We found that a necessary, but not sufficient, condition for the $\langle j \rangle=0$ macrostate to exist is that $P_{\lambda <0 } < 1/N$. Therefore, a sufficient, but not necessary, condition for the system to exist in a glassy regime is for this inequality to be violated.

Looking beyond this result, we might expect the introduction of disorder into our permutation system to come with the multiple equilibria and ultrametricity of SK spin glasses \cite{binder1986spin}. However,  for the class of permutation glasses considered here, this is not the case. We can see this by computing the Landau free energy for this disordered system. By \refew{free_en1_perm}, \refew{distr1}, and the substitution $s-1 \to j$, we find
\begin{equation}
\beta {\cal F}(j)= 1+j  -N\int^{\infty}_{-\infty} d\lambda \, \rho_{0}(\lambda) \ln \Big(1 + j e^{-\beta \lambda}\Big),
\label{eq:free_en_perm}
\end{equation}
In \reffig{phases}, we schematically plot this free energy, noting that it exhibits all of the functional forms of the free energy for the non-disordered case $\rho_{0}(\lambda) = \delta(\lambda- \lambda_0)$. Mathematically, this arises due to its stability conditions: Because the free energy for the glassy system is always convex, it can have at most one minimum and, by the constraints of this system, this minimum must occur somewhere in the range of  $0 \leq j \leq N$. However, not all forms of this free energy are accessible for all parameter values in the system. In particular, for the glassy regime in which \rfw{main_identity} has no solution, the free-energy curve with a global minimum at $\langle j \rangle=0$ cannot be achieved and the system has $\langle j \rangle\neq 0$ for all temperatures. 

Thus, for the simple permutation glass considered in this paper, we say the system exists in the "glassy regime" if and only if \rfw{main_identity} does not admit a solution for $\beta_c$ even when the mean energy costs $\overline{\lambda}$ is greater than zero. This simplest version of a permutation glass does not exhibit the replica symmetry breaking and ultrametricity characteristic of SK glasses, but we argue for the labeling of a particular regime as "glassy" due to its differing properties from the non-glassy regime: In the non-glassy regime, the disorder is not large enough to lead to thermal behavior different from that for the non-disordered system. However, in the glassy regime, the disorder is so large that at zero-temperature the system can exist in multiple microstates even if there is only a single free energy minima as a function of $j$.

\section{DISCUSSION \label{sec:five}}

Motivated by the importance of the orderings of amino acid sequences in the structure and function of proteins, a model was previously proposed to study the equilibrium thermodynamics of a system where particular permutations of an ordered list defined various energy states of the system. In that model, for simplicity and solubility, it was imposed that all lattice sites had the same energy cost for an incorrectly ordered component. However, more generally, it would have been useful to consider a system of permutations where the energy cost for each lattice site was drawn from a quenched distribution of energy costs.

We considered such permutation glasses here. The replica symmetric ansatz of such glasses yielded a result consistent with the thermodynamically stable state computed by applying Laplace's method to the partition function. We found that this simplest permutation glass does not exhibit the replica symmetry breaking of spin glasses. However, it does exhibit a glassy regime--characterized by $\langle j \rangle \neq 0$ for all temperatures--if \rfw{main_identity} cannot be satisfied even when $\overline{\lambda}>0$. In the non-glassy regime, the $\langle j \rangle=0$ state can be achieved but the transition temperature satisfies $k_BT_c \leq \overline{\lambda}/\ln N$, and thus the system is less tolerant of thermal disorder than is the non-disordered system in moving to the $\langle j \rangle=0$ state. 


From this analysis we found that we must have $P_{\lambda<0} <1/N$ in order for $\langle j \rangle=0$ to be a possible macrostate, that is, in order for the completely correct ordering to be an achievable thermal equilibrium and for the system to be in the non-glassy regime, the probability of having an energy benefit for an incorrectly ordered component must be less than the inverse of the number of components in the system. 

But having considered the permutation glass defined by the ``non-interacting" Hamiltonian \refew{ham_glass}, a natural extension would be to consider a permutation glass with the typical spin glass-like Hamiltonian 
\begin{equation}
{\cal H} = \sum_{i< j} \mu_{ij} I_{\theta_i \neq \omega_i} I_{\theta_j \neq \omega_j},
\label{eq:int_ham}
\end{equation}
where $\mu_{ij}$ is drawn from a distribution of interaction energies. Such a Hamiltonian associates an energy cost $\mu_{ij}$ with a permutation where both component $i$ and component $j$ are in an incorrect position. When the global analog of \refew{int_ham} was studied in \cite{williams2017statistical}, we found non-trivial regime behavior including multiple metastable states, multiple transition temperatures, and quadruple and triple points. Thus, considering the disordered behavior of a system with Hamiltonian \refew{int_ham}, should yield some novel results (such as replica symmetry breaking) over the simpler phase behavior depicted in \reffig{perm_glass_phase}.

Also, it is well known that spin glasses and other disordered systems often exhibit non-exponential relaxation behavior and memory effects \cite{binder1986spin}, thus an interesting question would be whether such properties exist in kinetic permutation glasses. Answering such a question would likely require studying glasses defined by \refew{int_ham} rather than \refew{ham_glass}. In spin glass models, the glass transition temperature is important in defining the onset of such non-exponential relaxations. However, the analog of such a temperature does not seem to exist in the model defined by \refew{ham_glass}. Thus, before a kinetic analysis of permutation glasses can yield additional insights into the non-equilibrium properties of disordered systems, it would likely prove necessary to consider more complex glass models than the one considered in this paper. 

Finally, we note that our free energy \refew{free_en_perm} is reminiscent of the thermodynamic potential of a familiar system in statistical mechanics. For a fermion system with a countably finite (but large) number of energy levels $N_{\text{lvl}}$ where each level is labeled $\ve_{k}$ for some integer $k$, the grand canonical potential of the system is \cite{landau1980statistical}
\begin{align}
\beta \Omega_{\text{Fermi}} & = - \sum_{k} \ln \left(1+  e^{\beta(\mu- \ve_k)}\right)\mm
& = - N_{\text{lvl}}\int^{\infty}_{-\infty} d\ve\, g(\ve)\, \ln\left(1+ e^{\beta(\mu- \ve)}\right)
\label{eq:fermi}
\end{align}
where $\mu$ is the chemical potential,  $g(\ve)$ is an energy density, and we used the heuristic \refew{distr1} to replace the discrete sum with an integral. Comparing \refew{free_en_perm} for $j = \langle j \rangle$ to \refew{fermi}, we see that we can transform the former into the latter by making the substitutions $\beta F- 1- \langle j \rangle \to \beta \Omega_{\text{Fermi}}$, $\rho_0(\lambda) \to g(\ve)$, $N \to N_{\text{lvl.}}$, and $\langle j \rangle \to e^{\beta \mu}$.

With these substitutions, we find that for the fermion system the condition analogous to \refew{lami_3} is 
\begin{equation}
e^{\beta \mu}= N_{\text{lvl.}}\int^{\infty}_{-\infty} d\ve  \frac{g(\ve)}{e^{\beta( \ve-\mu)} +1}.
\label{eq:fermion_gas_constr}
\end{equation}
We recall that for fermion gases in the grand canonical ensemble, the average number of fermions is given by 
\begin{equation}
\langle n_{\text{Fermi}} \rangle = N_{\text{lvl.}}\int^{\infty}_{-\infty} d\ve  \frac{g(\ve)}{e^{\beta (\ve-\mu)} + 1},
\label{eq:fermi_gas}
\end{equation}
a result which is reproduced by \refew{fermion_gas_constr} if we use the additional grand canonical ensemble constraint $\langle n_{\text{Fermi}} \rangle = e^{\beta \mu}$. 

Given the transformation $\langle j\rangle \to e^{\beta \mu}$, we then see that \refew{fermion_gas_constr} and \refew{fermi_gas} imply that in translating from a permutation glass to a fermion gas, we should interpret the order parameter $\langle j \rangle$ as the average number of fermions $\langle n_{\text{Fermi}} \rangle$. Therefore, the permutation glass inequalities $0\leq \langle j \rangle/N\leq1$ correctly imply the fermion gas inequalities $0\leq\langle n_{\text{Fermi}} \rangle/N_{\text{lvl.}}\leq 1$. Thus, the canonical ensemble of a simple permutation glass seems to be dual to the grand canonical ensemble of a fermion system with a large number of energy levels and where the chemical potential is given by $\beta \mu = \ln\langle n_{\text{Fermi}} \rangle$. 

Perhaps such a correspondence is not so surprising since permutations are central to the formalisms of both systems. Still, it is worth asking whether this duality can allow the understanding of one system to yield insights into the other. 

\begin{acknowledgments}
The author thanks Eugene Shakhnovich for comments and advice on an early draft of this paper, and Rostam Razban, and Evgeny Serbryany for discussions on the model contexts to which the system can be applied. He also thanks Michael Gaichenkov for questions concerning the clarity of various figures.
\end{acknowledgments}

\section{Supplementary Code \label{sec:supp}}
The code for creating \reffig{Tc_disorder}, \reffig{simulation2}, and \reffig{logprob_plot} can be found at: \href{https://github.com/mowillia/permutation_code}{https://github.com/mowillia/permutation\_code}.

\appendix

\section{Derivation of Correlation \label{app2}}
For our permutation system with the partition function 
\begin{equation}
Z_{N}(\beta\lambda_0) = \sum_{\{\vec{\theta}\}}\exp\left(-\beta\lambda_0 \sum_{i=1}^N I_{\theta_i \neq \omega_i}\right),
\label{eq:sim_part}
\end{equation}
the sum of all the site-site correlations is given by 
\begin{equation}
 \sum_{i,j}^N \sigma_{ij}^2 = \sum_{i, j}^N \Big( \langle I_{\theta_i \neq \omega_i} I_{\theta_j\neq \omega_j} \rangle - \langle I_{\theta_i \neq \omega_i} \rangle \langle I_{\theta_j\neq \omega_j} \rangle \Big).
\end{equation}
Given that no site is special we can expect the the site-site correlations for different sites to be the same regardless of which two sites we choose. Thus, we have
\begin{align}
\sum_{i,j}^N \sigma_{ij}^2 
& =  N(N-1)\sigma_{i\neq j}^2+\sum_{i=1}^N \Big( \langle I_{\theta_i \neq \omega_i} \rangle - \langle I_{\theta_i \neq \omega_i} \rangle^2 \Big),
\end{align}
where we used $I_{\theta_i \neq \omega_i}^2 = I_{\theta_i \neq \omega_i}$ in the last line. Thus we find that the site-site correlation for different sites is
\begin{align}
\sigma_{i \neq j}^2 &= \frac{1}{N(N-1)}\left[ \frac{\partial^2}{\partial(\beta\lambda_0)^2}\ln Z_{N}(\beta\lambda_0) \right. \mm
& \left. \hspace{2cm}- \sum_{i=1}^N \langle I_{\theta_i \neq \omega_i} \rangle\Big(1  - \langle I_{\theta_i \neq \omega_i} \rangle \Big)\right].
\label{eq:sigmi}
\end{align}
From \cite{williams2017statistical}, we have 
\begin{equation}
\ln Z_{N}(\beta\lambda_0) \simeq - N \beta\lambda_0 + e^{\beta\lambda_0} - N - 1 + G_{0}(N),
\label{eq:app_part}
\end{equation}
where $G_0(N)$ is independent of $\beta\lambda_0$. We also have that average incorrectness of a single site is 
\begin{equation}
\langle I_{\theta_i \neq \omega_i} \rangle \simeq 1  -e^{\beta\lambda_0}/N. 
\label{eq:incorr}
\end{equation}
Using \refew{incorr} and \refew{app_part} in \refew{sigmi} we obtain
\begin{equation}
\sigma_{i\neq j}^2 \simeq \frac{1}{N-1} \left( \frac{e^{\beta\lambda_0}}{N}\right)^2, 
\label{eq:sigm_res}
\end{equation}
which, given the limits of the Laplace's method result \refew{incorr},  is only valid for $\beta\lambda_0 < \ln N$.

\begin{widetext}
\section{Replica Symmetric Solution \label{rsym}}
In this appendix, we show \refew{lami_3} is consistent with the replica symmetric solution to the permutation model with quenched disorder.  To study quenched disorder in our permutation system, we must evaluate the quantity
\begin{equation}
\left\langle \ln Z_{N}(\{\beta \lambda_i\}) \right \rangle = \int^{\infty}_{-\infty}\prod_{k=1}^{N} d\lambda_k\, \rho(\{\lambda_j\}) \ln  \int^{\infty}_{0} dt\, e^{-s} \prod^{N}_{\ell=1}\Big( 1+ (s-1)e^{-\beta\lambda_{\ell}}\Big).
\end{equation}
For generality we will not specify a particular form for $\rho(\{\lambda_k\})$ other than to assume each $\lambda_k$ has the same distribution: 
\begin{equation}
\rho(\{\lambda_k\}) = \prod_{j= 1}^N \rho_{0}(\lambda_j).
\end{equation}
To implement the replica procedure, we apply the identity 
\begin{equation}
\ln Z = \lim_{n\to 0} \frac{Z^n-1}{n}, 
\label{eq:repl_id}
\end{equation}
and then compute $\langle Z^n \rangle $. Doing so, given the definition of $Z$ and our distribution of $\lambda_k$ values, we have
\begin{align}
\left\langle Z_{N}(\{\beta \lambda_i\})^n \right \rangle & = \int^{\infty}_{-\infty} \prod_{k=1}^{N} d\lambda_k\,\rho_{0}(\lambda_k)\,\int^{\infty}_{0} \prod_{\beta = 1}^n ds_\beta\, \, e^{-\sum_{\alpha=1}^n s_{\alpha}} \prod_{i=1}^N \prod_{\alpha=1}^n \left( 1+ (s_{\alpha}-1) e^{-\beta\lambda_i}\right),
\label{eq:repl_symm}
\end{align}
where Greek indices denote our replicas while Roman indices denote lattice places. 

Now, to make progress, we posit a replica symmetric ansatz in place of \refew{repl_symm}. The motivation for this replacement is that we introduced our replicas as an analytic trick, and they are thus unphysical aspects of our analysis. Therefore, any distinguishing elements between two replicas are unphysical. In the absence of any other supporting evidence, this motivation is in general an insufficient reason to accept the replica symmetric solution as valid, but we will find that this solution reproduces the thermodynamically stable result \refew{lami2} which was derived through alternative means. 

For the replica symmetric ansatz, we replace our distinct $n$ replica variables $s_1, \ldots, s_n$ with the single variable $s$. Doing so, we obtain
\begin{align}
\left\langle Z_{N}(\{\beta \lambda_i\})^n \right \rangle &\to  \int^{\infty}_{-\infty} \prod_{k=1}^{N} d\lambda_k\, \rho_{0}(\lambda_k) \int^{\infty}_{0} ds \, e^{-n  s} \prod_{i=1}^N  \left( 1+ (s_{\alpha}-1) e^{-\beta\lambda_i}\right)^n \mm
& = \int^{\infty}_{0} ds\, \exp \left[ - n s + \ln \Tr\, \exp L_{n}\big(s, \{\lambda_k\}\big) \right],
\label{eq:ansatz}
\end{align}
 where we defined
  \begin{align}
   \Tr[\cdots]  & \equiv \int^{\infty}_{-\infty}\prod_{k=1}^{N} d\lambda_k \, [\cdots] \\
L_{n}\big(s, \{\lambda_k\}\big)  & \equiv   \sum_{k=1}^{N} \ln \rho_{0}(\lambda_k)  + n \sum_{k=1}^N \ln \left( 1+ (s-1)e^{-\beta\lambda_k}\right).
  \end{align}
 Computing \refew{ansatz} via Laplace's method, and using the identity \refew{repl_id}, we find the quenched average free energy to be 
\begin{align}
\langle \ln Z \rangle & = \lim_{n\to 0} \frac{1}{n} \left\{\exp\Big[-ns_0 + \ln \Tr \exp L_{n}(s_0, \{\lambda_k\})\Big]- 1\right\}\mm
& = - s_0 + \lim_{n\to 0} \frac{1}{n} \ln \Tr \exp L_{n}(s_0, \{\lambda_k\}),
\end{align}
where $s_0$ is defined by the condition 
\begin{equation}
- 1 + \lim _{n \to 0}\frac{1}{n}\frac{\partial}{\partial s}  \ln \Tr \exp L_{n}(s, \{\lambda_k\}) \Big|_{s = s_0} = 0.
\label{eq:ton}
\end{equation}
Computing the argument of the limit in \refew{ton}, we find 
\begin{align}
\frac{\partial}{\partial s} \Tr \exp L_{n}\big(s, \{\lambda_k\}\big) & = \frac{\partial}{\partial s}  \int^{\infty}_{-\infty}\prod_{k=1}^{N} d\lambda_k\,\exp\left[ \sum_{k=1}^{N} \ln \rho_{0}(\lambda_k)+ n \sum_{k=1}^N \ln \left( 1+ (s-1)e^{-\beta\lambda_k}\right)\right]\mm
& = \int^{\infty}_{-\infty}\prod_{k=1}^{N} d\lambda_k\,\exp\left[L_{n}\big(s, \{\lambda_k\}\big)\right] \sum_{k=1}^N \frac{n e^{-\beta\lambda_{k}}}{1 + (s-1)e^{-\beta\lambda_k}}.
\end{align}
Thus given \refew{ton}, we have that $s_0$ must satisfy
\begin{align}
1 & =   \lim _{n \to 0}\int^{\infty}_{-\infty}\prod_{k=1}^{N} d\lambda_k \,\exp\left[L_{n}\big(s_0, \{\lambda_k\}\big)\right] \sum_{k=1}^N \frac{ e^{-\beta\lambda_{k}}}{1 + (s_0-1)e^{-\beta\lambda_k}}\mm
& = \int^{\infty}_{-\infty}\prod_{k=1}^{N} d\lambda_{k} \, \rho_{0}(\lambda_{k}) \sum_{k=1}^N \frac{ 1}{e^{\beta\lambda_k}+{s}_0-1} \mm
& = N\int^{\infty}_{-\infty} d\lambda \, \frac{\rho_{0}(\lambda)}{e^{\beta\lambda}+s_0-1}, 
\label{eq:lami_4}
\end{align}
where we used the independent normalization of each $\rho_{0}(\lambda)$ in the final line. Given the definition $s_0 - 1 = \langle j \rangle$, \refew{lami_4} is identical to \refew{lami_3}. The consistency between the replica symmetric ansatz and \refew{lami_3} suggests that this system of quenched disorder does not bear the more interesting features (e.g., multiple equilibria and ergodicity breaking) of replica symmetry breaking solutions to statistical mechanics systems. 
\end{widetext}

\section{Heuristic Derivation of \refew{tc} \label{app:heuristic}}
We derive \refew{tc} heuristically and thus lend quantitative justification to the qualitative argument outlined in \refsec{three}. We begin with the simple permutation model with no disorder. The energy of a microstate in such a system is $E =\lambda_0 j$ where $\lambda_0$ is the energy cost of an incorrect component and $j$ is the number of incorrect components. Also, the number of such microstates for a given $j$ is $\binom{N}{j} d_{j}$ where $N$ is the number of components in the system, and  $d_j$ is the number of derangements of a list with $j$ elements. Thus, the microcanonical ensemble entropy for a given $E$, $\lambda_0$, and $N$ is 
\begin{align}
S_N(E, \lambda_0) & = k _B \ln \left[ \binom{N}{E/\lambda_0} d_{E/\lambda_0}\right] \mm
& \simeq - k_B \ln \Gamma(N- E/\lambda_0 +1) + k _B \ln \Gamma(N+1).
\label{eq:sne}
\end{align}
If we were to introduce a small amount of disorder $\sigma_0$ into our system, such that $\lambda_0$ (instead of being fixed at a single value) had a non-negligible probability to be found within the domain $[\lambda_0 - \sigma_0, \lambda_0+\sigma_0]$, then we could approximate this new entropy as a two-point average over the ends of this domain. Defining this entropy as $\langle S(E) \rangle_{\lambda_0, \sigma_0}$ we have
\begin{align}
\langle &S_N(E, \lambda_0)   \rangle_{ \sigma_0} \mm
& \equiv \frac{1}{2}S_N(E, \lambda_0-\sigma_0)  + \frac{1}{2}S_N(E, \lambda_0+ \sigma_0) \mm
& = S_N(E, \lambda_0) + \frac{\sigma_0^2}{2} \frac{\partial^2}{\partial \lambda^2} S_N(E, \lambda)\Big|_{\lambda=\lambda_0}+ {\cal O}(\sigma_0^4). 
\label{eq:dis_sne}
\end{align}
We note that \refew{dis_sne}, given the convexity of $S(N,\lambda)$ with respect to $\lambda$, is consistent with the intuition that introducing disorder into our system effectively increases the entropy. By the thermodynamic definition, the temperature of this disordered system is 
\begin{align}
\frac{1}{T(\sigma_0)} & =  \frac{\partial}{\partial E}\langle S_N(E, \lambda_0) \rangle_{\sigma_0}.
\label{eq:crite}
\end{align}
Our goal is to compute the transition temperature for the $\langle j \rangle = 0$ transition. By $E= \lambda_0j$, we take this transition temperature to be the same as that associated with a microstate energy $E=0$ in \refew{crite}. Using \refew{sne}, we thus find
\begin{align}
\frac{1}{T_c(\sigma_0)} & = \frac{\partial}{\partial E}\langle S_N(E, \lambda_0) \rangle_{\sigma_0} \Big|_{E=0} \mm
& =  \frac{\partial}{\partial E} S_N(E, \lambda_0) \Big|_{E=0} + \frac{\sigma_0^2}{2} \frac{\partial}{\partial E}\left[\frac{\partial^2}{\partial \lambda^2} S_N(E, \lambda)\Big|_{\lambda=\lambda_0}\right]_{E=0}\mm
&\qquad+  {\cal O}(\sigma_0^4/\lambda_0^4)\mm
& = \frac{k_B \ln N}{\lambda_0} + \frac{2k_B}{\lambda_0^3} \cdot \frac{\sigma_0^2}{2} \ln N +  {\cal O}(\sigma_0^4/\lambda_0^4) ,
\label{eq:critt}
\end{align}
where we used $\psi_0(N) \simeq \ln(N)$ (with $\psi_0$ being the digamma function) for $N\gg 1$. \refew{critt} then implies
\begin{equation}
k_BT_c(\sigma_0) = \frac{\lambda_0}{\ln N} \left[ 1 - \frac{\sigma_0^2}{\lambda_0^2} + {\cal O}(\sigma_0^4/\lambda_0^4) \right],
\end{equation}
which reproduces, up to an order of magnitude, the ${\cal O}(\sigma_0^2)$ correction in \refew{tc}.

\begin{widetext}

\section{Order Parameter for The Symmetric Bernoulli distribution \label{app:bernoulli}}
We will use \refew{lami_3} to derive an exact expression for the order parameter of the permutation glass with a symmetric Bernoulli distribution of energy costs. There seem to be no clean analytic expressions for the order parameters associated with the Gaussian or uniform distributions of energy costs. 

Integrating the distribution \refew{distr1} according to \refew{lami_3}, we find
\begin{equation}
\frac{1}{N} =\frac{q}{e^{\beta\bar{\lambda}} + \langle j \rangle}+ \frac{1-q}{e^{-\beta\bar{\lambda}} + \langle j \rangle}.
\label{eq:quad_res}
\end{equation}
Solving \refew{quad_res} and dropping the solution which does not reduce to $N - e^{\beta\lambda}$ in the $q\to 0$ limit, we find the order parameter
\begin{align}
\langle j\rangle /N &= \frac{1}{2}\left[ 1 - \frac{2}{N} \cosh(\beta\bar{\lambda})+ \right.\left. \sqrt{1 + \frac{4}{N} (1- 2q)\sinh(\beta\bar{\lambda})+ \frac{4}{N^2} \sinh^2(\beta\bar{\lambda})} \,\right],
\label{eq:jd}
\end{align}
where $\langle j \rangle$ could be written in terms of $\lambda_0$ and $\sigma_0$ by inverting the system \refew{avg_sig}. 
\end{widetext}

\section{Deriving Probability limits \label{app:prob_distr}}
We derive the probabilities \refew{probf}, \refew{probg}, and \refew{probd} which establish the constraints the respective distributions must satisfy in order for $\beta_c$ to exist and $\langle j \rangle=0$ to be an equilibrium. We begin with the mean-variance inequalities  \refew{gauss_ineq}, \refew{flat_ineq3}, and \refew{dirac_ineq} expressed as limits on the maximum value of the standard deviation:
\begin{equation}
\sigma_ 0  \leq   \sigma_{0}^{\text{max} }  = \begin{dcases}  \frac{\lambda_0}{\sqrt{3}} \left( 1- \frac{2}{Ne}\right) ^{-1} &  \text{[Uniform]} \\
\frac{\lambda_0}{\sqrt{ 2\ln N}} & \text{[Gaussian]} \\
\frac{\lambda_0}{\sqrt{N^2-1}} &  \text{[Symm. Bernoulli]}
\end{dcases}
\end{equation}
In order for $\beta_c$ to exist, the mean $\lambda_0$ of each distribution must be greater than zero. Consequently the maximum values of $\sigma_0$ must be associated with maximum probabilities of obtaining a $\lambda<0$ from the distribution.  Computing these probability inequalities for each distribution, we find
\begin{align}
 P^{\text{uniform}}_{\lambda<0} & \leq \frac{1}{2\sigma_{0}^{\text{max} }\sqrt{3} }\left( \sigma_{0}^{\text{max}}\sqrt{3} - \lambda_0\right)\mm
&  = \frac{1}{Ne}\\
P^{\text{gauss}}_{\lambda<0} &\leq \int^{0}_{-\infty}\frac{d\lambda}{2\pi(\sigma^{\text{max}})^2_0} e^{-(\lambda-\lambda_0)^2/2(\sigma_{0}^{\text{max}})^2}\mm
& = \frac{1}{2}\left[1- \text{erf}\left(- \sqrt{\ln N}\right) \right]  \simeq \frac{1}{2 N \sqrt{\pi \ln N}}\\
P^{\text{bernoulli}}_{\lambda<0} & \leq \frac{1}{2}\left( 1- \sqrt{1- \frac{1}{N^2}}\right) \mm
& \simeq \frac{1}{4N^2}, \label{eq:dirac_prob}
\end{align}
Where each quantity is expanded in the large $N$ limit where relevant, and \refew{dirac_prob} follows from \refew{refq} and the equality $1- q = P^{\text{bernoulli}}_{\lambda<0}$.


\bibliographystyle{ieeetr}

\end{document}